# Arctic Oscillation Modulation of Winter Air–Sea Coupling in the East/Japan Sea: Persistence, Timescales, and Extremes

**Gyuchang Lim[1], JongJin Park[1,2]***

[1]Kyungpook Institute of Oceanography, Kyungpook National University, Daegu, Korea

[2]School of Earth System Sciences, Kyungpook National University, Daegu, Korea

**\* Correspondence:**
JongJin Park
jjpark@knu.ac.kr

**Abstract**

The winter climate of the East/Japan Sea (EJS) is strongly affected by the Arctic Oscillation (AO), yet how AO polarity reshapes the memory, coupling patterns, and predictability of sea-surface temperature anomalies (SSTA) remains poorly quantified. Using 30 winters (1993–2022) of daily OISST and ERA5 fields, we combine multivariate Maximum Covariance Analysis (MCA) with an Ornstein–Uhlenbeck (OU)-like integration of atmospheric principal components (PCs). The leading coupled mode explains 87% (+AO) and 75% (−AO) of squared covariance, with SSTA hot spots in East Korea Bay and along the subpolar front. Zero-lag correlations between the SSTA PC and OU-integrated atmospheric PCs reveal characteristic memory timescales ($\tau^*$) of ~18–25 days for wind-stress curl (CurlTau), ~15–30 days for near-surface air temperature (ATMP) and zonal winds, and ~30–50 days for sea-level pressure (SLP) and meridional winds—longer under −AO. Detrended Fluctuation Analysis (DFA) shows SSTA persistence $H \approx$ 1.3–1.4 and that integrated atmospheric responses acquire ocean-like persistence, validating Hasselmann's stochastic framework for winter EJS. AO-phase contrasts align with a curl→Ekman pumping→eddy/SSH→SST pathway: +AO favors anticyclonic/downwelling responses and warmer SSTA, whereas −AO favors cyclonic/upwelling and cooler SSTA. These diagnostics identify phase-specific predictor windows (e.g., 3-week OU-integrated CurlTau/ATMP; 4–7-week SLP/V-wind under −AO) to initialize subseasonal extremes prediction (marine heatwaves and cold-surge-impacted SST). The approach quantifies memory scales and spatial coupling that were not explicitly resolved by previous composite analyses, offering a tractable foundation for probabilistic forecast models.

## 1    Introduction

The East/Japan Sea (EJS)—a deep, semi-enclosed marginal sea bordered by Korea, Japan, and Russia—behaves as a miniature ocean, hosting western boundary currents, sharp fronts, deep winter mixing, and vigorous eddies within a compact basin. This dynamical richness imprints strong wintertime variability on sea-surface temperature and modulates extremes such as marine heatwaves (MHWs) and cold surges. Past studies have linked EJS variability to local wind forcing and air–sea heat exchange, as well as to larger-scale climate modes (East Asian winter monsoon, ENSO) (Park & Chung, 1999; Park & Oh, 2000; Minobe et al., 2004; Jeong et al., 2022). During boreal winter, the Arctic Oscillation (AO) has emerged as an additional, and increasingly important, driver over the Korea–Japan–Russia sector (Park & Chu, 2006; Thompson & Wallace, 2000; Lee et al., 2010; Gong et al., 2001; Zhang et al., 2014). Its positive phase (+AO) is typically associated with a weakened East Asian winter monsoon, while its negative phase (−AO) strengthens cold-air outbreaks and storminess (Gong et al., 2001; Park et al., 2011; Kim et al., 2022). Recent analyses further suggest that AO polarity projects onto wind-stress curl, Ekman pumping, and eddy/sea-surface height (SSH) anomalies that precondition the northwestern EJS (NW EJS) for warm or cold SST states and influence the likelihood of winter MHWs (Song et al., 2023; Hobday et al., 2016).

Despite this progress, two aspects of AO–EJS coupling remain poorly quantified. First, most composite or lag-regression analyses implicitly treat the atmosphere as an instantaneous driver of SST and therefore offer limited insight into the time-integration by which synoptic "weather" is accumulated by the ocean mixed layer. Second, the spatial structure of air–sea coupling—in particular, how AO polarity focuses forcing along fronts and boundary-current pathways—has rarely been mapped in a way that is both multivariate (across several atmospheric fields) and explicitly phase-conditioned on AO. These gaps hinder the identification of predictability windows for subseasonal extremes.

The present study is motivated by Hasselmann's stochastic climate framework, which conceptualizes the ocean mixed layer as a slow integrator of rapidly varying atmospheric noise, leading to reddened SST spectra and enhanced persistence at subseasonal time scales (Hasselmann, 1976; von Storch & Zwiers, 1999; Majda et al., 2001). In its simplest form with linear damping, the ocean's response can be represented by an Ornstein–Uhlenbeck (OU) process, characterized by an e-folding memory time scale $\tau$ that governs both persistence and predictability. We posit that AO polarity reorganizes this effective memory kernel—not just the mean state of the forcing. In particular, we expect the curl → Ekman pumping → eddy/SSH → SST pathway (Song et al., 2023) to operate on short time scales, whereas basin-scale pressure and meridional wind patterns exert longer conditioning, with relative dominance that depends on AO phase.

To address these issues, we combine three ingredients. First, we map wintertime persistence using Detrended Fluctuation Analysis (DFA) for SST and a suite of ERA5 atmospheric fields (Peng et al., 1994; Lim & Park, 2024a; Lim & Park, 2024b), thereby quantifying Hurst exponents (H) and diagnosing whether SST behaves as an integrated process at 10–90-day scales. Second, we extract the leading coupled mode between SST anomalies (SSTA) and the atmospheric fields using multivariate Maximum Covariance Analysis (MCA), explicitly stratified by AO phase (+AO, −AO) to reveal hot spots (e.g., East Korea Bay and the subpolar front (SPF)) and front-parallel wind-curl bands. Third, we introduce an OU-like integration of the atmospheric principal components (PCs) and optimize the e-folding memory time $\tau^*$ by maximizing the zero-lag correlation with the

SSTA-mode PC; we then verify, via DFA, that the integrated responses acquire ocean-like persistence, consistent with Hasselmann's hypothesis. Collectively, these diagnostics quantify memory scales and spatial coupling patterns under AO polarity and thereby identify phase-specific predictor windows relevant for the probabilistic prediction of winter extremes.

Our analysis uses daily OISST (1993–2022) and ERA5 fields for the EJS domain (34–45° N, 127–144° E). We isolate JFM ("cut-and-stitch") winters, tag each winter by its AO polarity, and examine persistence (DFA), coupled patterns (rank-reduced multivariate MCA with saliency masks and unmasked maps), and memory diagnostics (OU-like response integration across a 2–50-day $\tau$ grid). This design allows us to (i) compare AO-conditioned persistence in SST and atmospheric fields, (ii) isolate coupled hot spots and front-parallel curl structures, and (iii) estimate $\tau^*$ for each atmospheric driver and AO phase. Full details of datasets, AO stratification, rank reduction, saliency criteria, OU implementation, and correlation metrics are provided in Methods.

Guided by this framework, we hypothesize that the EJS mixed layer integrates rapidly varying atmospheric noise over characteristic memory windows whose length and efficacy depend on AO polarity. We therefore seek to estimate these memory windows objectively and to test whether AO-conditioned differences in pressure–wind–curl forcing alter the strength and persistence of the coupled mode.

By providing quantitative memory scales and spatial coupling templates that are explicitly conditioned on AO polarity, our work extends earlier composite studies that emphasized AO's sign and spatial imprint but did not resolve integration time scales or phase-dependent persistence. The results define operationally useful predictor windows and set the stage for phase-aware, OU-driven simulations to estimate the probability of winter extremes such as marine heatwaves and cold-surge-impacted SST anomalies in the EJS (Khelif et al., 2005).

This paper is organized as follows. Section 2 details data and methods (DFA, AO stratification, rank-reduced multivariate MCA, OU-like response integration). Section 3 presents persistence maps, coupled spatial patterns, and memory diagnostics, and contrasts +AO with −AO. Section 4 interprets the findings within a stochastic-dynamical framework and discusses implications for subseasonal predictability and extremes.

## 2 Methods

### 2.1 Data

#### 2.1.1 Sea Surface Temperature (SST)

We use the gridded daily Optimum Interpolation SST (OISST) version 2.1 product provided by the National Oceanic and Atmospheric Administration (NOAA; Huang et al. 2021; Reynolds et al. 2007). These data have a spatial resolution of 0.25° × 0.25°, spanning 1981–present, and combine observations from multiple platforms (satellites, ships, buoys, Argo floats) into a regularly gridded global dataset. For this study, we restrict our analysis to the East/Japan Sea (34°–45° N, 127°–144° E) over the period 1 January 1993 to 31 December 2022, ensuring consistency with previous regional investigations (Lim & Park 2024a; Lim &

Park 2024b). Also, the SST anomalies were obtained by removing the daily climatology defined over the full duration (Jan. 1993 to Dec. 2022).

### 2.1.2 Arctic Oscillation (AO) Phase Classification

The monthly AO index is obtained from the NOAA Climate Prediction Center (National Oceanic and Atmospheric Administration Climate Prediction Center, 2023). It is derived by projecting daily 1000 hPa height anomalies (from NCEP–NCAR reanalysis) onto the AO's leading mode of variability. Positive (negative) AO phases are characterized by relatively low (high) sea-level pressure over the Arctic and high (low) pressure in the mid-latitudes. We average daily AO values to form monthly means and normalize them according to the 1981–2010 climatology. For winter (January–February–March, JFM) analyses, we define positive AO years as those in which the seasonal AO mean exceeds +0.8 standard deviations, and negative AO years as those below −0.8 standard deviation [10].

### 2.1.3 Atmospheric and Oceanic Variables (ERA5)

We employ the ERA5 reanalysis dataset from the European Center for Medium-Range Weather Forecasts (ECMWF) to analyze atmospheric and oceanic drivers influencing SSTA persistence (Hersbach et al. 2020). ERA5 provides hourly and monthly data at a 0.25° × 0.25° resolution from 1940–present, assimilating observations from satellites, radiosondes, and surface stations into a comprehensive global model. All the ERA5 variables were retrieved at 6–hour intervals (00, 06, 12, and 18 UTC), averaged to daily means, and bilinearly re-gridded onto the OISST grid. In this study, we considered six atmospheric fields as follows:

(1) 2m Atmospheric Temperature (ATMP; $T_{2m}$) indicates temperature at 2 meters above the surface, reflecting near-surface atmospheric conditions affecting SST.

(2) Sea Level Pressure (SLP; $P_{SL}$) indicates atmospheric pressure patterns (hPa), critical for identifying synoptic scale systems such as the AO.

(3) U and V wind components at 10m (U10, V10; $u_{10}, v_{10}$) are separately considered in this study.

(4) Wind Stress Curl (CurlTau; $\nabla \times \tau$) were computed by applying a numerical central difference method to the wind stress fields obtained from a pair of U and V wind components at every grid. The WSC is known to influence SST variability via ocean mixing and heat flux.

### 2.1.4 Wintertime Subsampling

To focus on boreal winter seasons of all datasets, we extract only the JFM portion from the full daily time series of each field by applying a cut-and-stitch procedure: the JFM segments from each year are concatenated in chronological order to form a single "wintertime" SSTA and atmospheric time series. Then, each subsampled time series is further divided into two sub-series tagged by its AO phase (positive and negative). Finally, we obtain two AO-tagged datasets for SSTA and atmospheric fields.

## 2.2 Analysis Methodology

### 2.2.1 Detrended Fluctuation Analysis (DFA)

We apply the conventional DFA algorithm (Peng et al. 1994) to each AO-phase tagged time series. The procedure is as follows:

**Signal Profile Construction**: Let $\{x_i\}$ for $i = 1,2,\cdots,N$ be a wintertime AO-phase tagged series. Define the signal profile,

$$X(k) = \sum_{i=1}^{k}[x_i - \langle x \rangle] \tag{1}$$

where $\langle x \rangle$ is the mean of $\{x_i\}$ over the full duration.

**Segmenting the Profile**: Choose a range of segment scales $s$ (herein, $10 \leq s \leq 90$ days) appropriate for short-term winter analysis. For each $s$, divide $X(k)$ into $N_s = \text{int}(N/s)$ disjoint segments from the front, and an equal number of the rear, yielding $2N_s$ segments. Thus, for each segment $v$ ($v = 1,\cdots,2N_s$), we have a pair of segmented time series $\{x_v(k), X_v(k)\}$ for $k = 1,\cdots,s$.

**Detrended Variance Computation**: For each segment $v$, we remove the local $m$-th order polynomial fit $P_v^m$ from $X_v(k)$ and compute,

$$F_X^2(v,s) = \frac{1}{s}\sum_{k=1}^{s}\{|X_v(k) - P_v^m(k)| \times |X_v(k) - P_v^m(k)|\} \tag{2}$$

We average $F_X^2(v,s)$ over all segments to obtain a detrended fluctuation function,

$$F_X^2(s) = \left(\frac{1}{2N_s}\sum_{v=1}^{2N_s} F_X^2(v,s)\right) \tag{3}$$

**Scaling-invariance and Hurst Exponent Estimation**: If $F_X(s)$ scales as a power law in $s$, we write

$$F_X(s) \sim s^{h_X} \tag{4}$$

where $h_X$ is Hurst exponent, an indicator of persistence of a time series. In our wintertime, we fit the linear slope in the log–log plot of $F_X(s)$ vs. $s$ over the chosen range $5 \leq s \leq 45$ days. Hurst exponent $0.5 < h_X < 1$ indicates persistent behavior of a time series, with $0 < h_X < 0.5$ suggesting anti-persistence and $h_X \approx 0.5$ indicating a white noise. Notably, $h_X > 1.0$ generally implies a random-walk, an integration of a (fractional) random noise with

Hurst exponent being less than 1.0, which is of particular interest in this study; we assume that the oceanic field, SSTA, is an integration of white-noise like atmospheric fields.

### 2.2.2 Multivariate Maximum Covariance Analysis (MCA)

All anomaly fields are normalized by their JFM standard deviation and weighted by $\sqrt{\cos\varphi}$ to compensate for the meridional convergence of grid boxes (Bretherton et al. 1992). In order to analyze the covariance structure between SST anomaly ($\boldsymbol{X}$) and concatenated atmospheric fields anomaly ($\boldsymbol{Y}$), let

$$\boldsymbol{X} \in \mathbb{R}^{T\times M}, \quad \boldsymbol{Y} = [\boldsymbol{T}_{2m}, \boldsymbol{\nabla}\times\boldsymbol{\tau}, \boldsymbol{P}_{SL}, \boldsymbol{u}_{10}, \boldsymbol{v}_{10}] \in \mathbb{R}^{T\times N} \tag{5}$$

where each row is a winter day and each column indicates a grid point of the respective anomaly field.

The (sample) cross-covariance matrix

$$\boldsymbol{C} = \frac{1}{T-1}\boldsymbol{X}^\mathrm{T}\boldsymbol{Y} \tag{6}$$

is decomposed by singular-value decomposition (SVD)

$$\boldsymbol{C} = \boldsymbol{U}\boldsymbol{\Sigma}\boldsymbol{V}^\mathrm{T} \tag{7}$$

yielding paired spatial patterns $\boldsymbol{U}$ (SST anomaly) and $\boldsymbol{V}$ (concatenated atmospheric fields anomaly) along with singular values $\sigma_k$ on the diagonal of $\boldsymbol{\Sigma}$. The *squared covariance fraction* (SCF) of mode $k$,

$$\mathrm{SCF}_k = \frac{\sigma_k^2}{\sum_j \sigma_j^2} \tag{8}$$

quantifies the relative contribution of that mode to total inter-field covariance (Bretherton et al., 1992; Björnsson & Venegas 1997; DelSole & Tippett 2020).

**2.2.2.1 Sampling-error issue for $T \ll N$**

In our AO-stratified composites, the number of independent winter days is modest (AO+: T = 721; AO−: T = 542), whereas the concatenated atmospheric field contains N ≈ 7365 grid columns. When the ratio T/N ≪ 1, the sample cross-covariance $\boldsymbol{C}$ has rank ≤ T−1 and its leading singular vectors are easily contaminated by noise; Monte-Carlo tests show that, for T/N < 0.1, spurious modes can explain > 30 % of the apparent squared covariance (Bretherton et al. 1992).

To mitigate this bias, we follow the rank-reduction strategy detailed in Björnsson & Venegas (1997) and widely used in subsequent climate studies; the so-called EOF-truncation pre-filter is given as

**EOF reduction of each field**: For the SST anomaly matrix $X$, compute the eigen-decomposition

$$X = U_x S_x V_x^T \tag{9}$$

and retain the first $n_x$ columns (modes) so that a chosen fraction $\gamma$ of total variance is preserved:

$$\gamma = \frac{\sum_{i=1}^{n_x} S_{x,ii}^2}{\sum_j S_{x,jj}^2} \tag{10}$$

with $\gamma = 0.90$ (90% variance; adopted in this study). The reduced SST anomaly matrix is

$$\widetilde{X} = U_x(:,1:n_x) S_x(:,1:n_x) \in \mathbb{R}^{T \times n_x} \tag{11}$$

Each atmospheric field anomaly $Y^{(m)}$ is treated identically, keeping $n_y^{(m)}$ EOFs. Concatenation of the reduced PCs gives

$$\widetilde{Y} \in \mathbb{R}^{T \times n_y} \tag{12}$$

with $n_y = \sum_m n_y^{(m)} \ll N$.

**Rank-reduced cross-covariance**: The reduced cross-covariance

$$\widetilde{C} = \frac{1}{T-1} \widetilde{X}^T \widetilde{Y} \in \mathbb{R}^{n_x \times n_y} \tag{13}$$

is then a well-conditioned ($n_x \times n_y$) matrix with $T/n_y \gtrsim 5$, yielding the well-determined leading modes (rule-of-thumb in Bretherton et al. 1992).

**SVD and Back-projection**: The SVD of rank-reduced cross-covariance matrix (Eq. 12)

$$\widetilde{C} = \widetilde{U} \Sigma \widetilde{V}^T \tag{14}$$

produces singular values $\sigma_k$ and vectors $\widetilde{U}_{(:,k)}$, $\widetilde{V}_{(:,k)}$. The full-grid spatial patterns are obtained by the following back-projection

$$\boldsymbol{u}_k = \boldsymbol{V}_x(:,1:n_x)\widetilde{\boldsymbol{U}}_{(:,k)}, \quad \boldsymbol{v}_k = \left[\boldsymbol{V}_y^{(1)} | \cdots | \boldsymbol{V}_y^{(5)}\right]\widetilde{\boldsymbol{V}}_{(:,k)}. \tag{15}$$

This EOF-MCA approach reduces the effective atmospheric dimension from $N \approx 7{,}400$ to $n_y \approx 90$, raising the sample ratio to $T/n_y \approx 8.0$ (AO+) and 6.0 (AO−), which are well above the threshold recommended by Bretherton et al. (1992); $n_x = 23$. The leading mode's SCF (~0.87 for AO+; ~0.75 for AO−) match those from the unreduced matrix without the spurious rank−1 collapse seen when no truncation is applied.

### 2.2.3 Saliency mask for MCA 1st-mode spatial loadings

To emphasize robust spatial features, we constructed a saliency mask for the MCA 1st-mode loadings of SST anomaly ($u_1$) and for each atmospheric field anomalies ($v_1^{(j)}, j \in \{T_{2m}, \nabla \times \boldsymbol{\tau}, P_{SL}, \boldsymbol{u}_{10}, \boldsymbol{v}_{10}\}$). After normalizing each loading vector to unit maximum amplitude, a grid cell was deemed salient if its absolute loading exceeded the area-weighted mean absolute loading. The resulting masks highlight East Korea Bay (EKB) and the Sub-Polar Front (SPF) as persistent hot spots across AO phases. Because this is an effect-size filter rather than a formal significance test, the unmasked (original) loadings are provided in the Supplementary Materials for reference. The procedures are given as

**Normalization:** Each loading vector is rescaled to remove arbitrary column-wise amplitudes and enable comparison across AO phases and domains. We use max-absolute normalization

$$\hat{\boldsymbol{u}}_1 = \frac{\boldsymbol{u}_1}{\|\boldsymbol{u}_1\|_\infty}, \quad \hat{v}_1^{(j)} = \frac{v_1^{(j)}}{\left\|v_1^{(j)}\right\|_\infty} \tag{16}$$

so that $\max_i |\hat{u}_{1,i}| = \max_i |\hat{v}_{1,i}^{(j)}| = 1$. For plotting, we use a fixed color range $[-1,1]$.

**Salience mask by area-weighted threshold:** Let $w_i = \cos\varphi_i$ denote the usual latitude-based area weight at grid cell $i$. The area-weighted mean absolute loading is

$$\theta_u = \frac{\sum_i w_i |\hat{u}_{1,i}|}{\sum_i w_i}, \quad \theta_v^{(j)} = \frac{\sum_i w_i |\hat{v}_{1,i}^{(j)}|}{\sum_i w_i}. \tag{17}$$

We define the salient sets

$$S_u = \{i: |\hat{u}_{1,i}| \geq \theta_u\}, \quad S_v^{(j)} = \{i: |\hat{v}_{1,i}^{(j)}| \geq \theta_v^{(j)}\}. \tag{18}$$

Only grid cells in $S_u$ (or $S_v^{(j)}$) are displayed in the spatial loading maps; all others are masked.

### 2.2.4 Ornstein-Uhlenbeck (OU)-based integrated atmospheric response

To quantify how much recent atmospheric variability contributes to the leading SST anomaly mode, we transform each atmospheric principal component (PC) into an integrated response with an exponential memory of length $\tau$ days, then measure how strongly this response co-varies with the SST anomaly PC. The $\tau$ that maximizes the correlation is interpreted as the characteristic memory time-scale of the air-sea interaction in the EJS during winter.

#### 2.2.4.1 Model and kernel

Let $a_1(t)$ denote the first-mode SST anomaly PC from the EOF-MCA, and let $b_j(t)$ be the paired atmospheric PC for field $j \in \{\boldsymbol{T}_{2m}, \boldsymbol{\nabla} \times \boldsymbol{\tau}, \boldsymbol{P}_{SL}, \boldsymbol{u}_{10}, \boldsymbol{v}_{10}\}$. We assume that the SST-relevant atmospheric influence enters a linear OU response:

$$\frac{dr_{\tau,j}}{dt} = -\frac{1}{\tau} r_{\tau,j}(t) + b_j(t) + \sigma dW_t \tag{19}$$

In our study, the stochastic innovation term is omitted (we set $\sigma = 0$) so that the response is entirely determined by the observed $b_j(t)$. This choice makes the optimal integration time $\tau^*$ identifiable directly from the $\tau$-dependence of the cross-correlation $\rho_j(\tau)$ between $a_1(t)$ and $r_{\tau,j}(t)$; avoids conflating unexplained noise with the integration signal; and yields a conservative estimate of the coupling memory (adding noise would only dilute $|\rho_j(\tau)|$ without shifting $\tau^*$ in a stable system.

With the stationary solution (zero as $t \to -\infty$), the continuous-time solution becomes the causal exponential convolution

$$r_{\tau,j}(t) = \int_0^\infty \exp\left(-\frac{s}{\tau}\right) b_j(t-s) ds. \tag{20}$$

Eq.16 is the Green-function (impulse response) of Eq.15; it is equivalent to an exponential moving integral with e-folding time $\tau$.

#### 2.2.4.2 Discrete implementation

All time-series are daily and limited to winter (JFM). With $\Delta t = 1$ day, the causal exponential convolution (Eq.19) is implemented as

$$r_{\tau,j}(t) = \sum_{k=0}^{K-1} \exp\left(-\frac{k\Delta t}{\tau}\right) b_j(t-k)\Delta t, \qquad \tau \in [2, 50] \text{ days}. \tag{21}$$

Because we use Pearson correlation, absolute scaling is immaterial; including an explicit $\Delta t$ factor would only rescale $r_{\tau,j}$ without affecting $\rho_j(\tau)$.

**2.2.4.3 Correlation-based identification of the memory time-scale**

For each atmospheric field $j$ and each $\tau \in \{2, 3, \cdots, 50\}$ days, we compute the Pearson cross-correlation

$$\rho_j(\tau) = \text{corr}(a_1(t), r_{\tau,j}(t)), \quad \tau_j^* = \arg \max_{\tau \in [2,50]} |\rho_j(\tau)| \tag{22}$$

where $\tau_j^*$ denotes the optimal integration time. When determining $\tau_j^*$, the absolute value $|\rho_j(\tau)|$ is used because MCA eigenvectors are sign-ambiguous; the physics resides in the magnitude and the time-scale of coupling, not in an arbitrary sign choice. Further, we report $\tau_j^*$ together with the plateau-onset time $\tau_{s,j}$ defined as the smallest $\tau$ for which $|\rho_j(\tau)|$ first exceeds 95% of its maximum; this approach is useful when the curves are flat.

**2.2.4.4 Interpretation**

**Characteristic time-scale:** $\tau_j^*$ (or $\tau_{s,j}$ for plateau cases) is interpreted as the effective memory over which the mixed layer accumulates forcing from field $j$ to produce the leading SST anomaly mode. A broad plateau indicates a low-pass, near-integrator behavior; a dome-shaped peak indicates that integration beyond $\tau_j^*$ blends signal with oppositely signed variability (over-integration), reducing coherence with the SST anomaly PC.

**Sign of $\rho$:** Because MCA PCs are defined up to an overall sign, cross-correlation sign is not uniquely meaningful unless all PCs externally anchored. We therefore compare $|\rho|$ and the associated time-scales across AO phases and domains.

**Atmospheric Fields analyzed:** We apply Eqs. (21–22) to atmospheric fields anomaly PCs, with a common $\tau$ grid of 2–50 days.

# 3     Results

Our goal is to characterize how the AO phase conditions wintertime SSTA dynamics in the EJS by quantifying (i) persistence of SST and atmospheric anomalies, (ii) spatial co-variability between SSTA and multiple atmospheric fields, and (iii) the effective memory with which the mixed layer integrates atmospheric fluctuations. Guided by Hasselmann's framework, we expect winter SSTA to behave as an integrated (reddened) response to atmospheric "weather", with AO polarity reorganizing both the spatial coupling and the time scale of integration. To capture local and remote (synoptic-scale) couplings simultaneously, we combine DFA-based persistence diagnostics with a multivariate MCA and an OU-like response model for atmospheric principal components (PCs).

## 3.1   Persistence Characteristics

### 3.1.1 Wintertime Persistence

We first estimate Hurst exponents ($H$) for JFM SSTA and atmospheric anomalies over 10–90-day scales; a range that shows no wintertime crossover in our data. Earlier work on the full-year series over 1993–2023 reported a crossover near 200–300 days (Lim & Park 2024a; Lim & Park 2024b), from strongly persistent behavior ($H$ >1) on sub-annual scales toward weaker persistence/white-noise features at multi-year scales; restricting to JFM eliminates that crossover and isolates subseasonal memory.

Figure 1 maps SSTA persistence separately for +AO and −AO winters. With the exception of a narrow band near the Japanese coast, most of the basin exhibits $H>1$, and a patchy high-$H$ corridor extends from EKB toward the subpolar front (SPF). In +AO winters the NW hot spot is particularly persistent ($H \gtrsim 1.5$), consistent with conditions that favor frequent warm anomalies and winter marine heatwaves in that sector (Song et al. 2023; Hobday et al. 2016).

FIGURE 1 TO BE INSERTED

Figure 2 shows persistence for five atmospheric fields. All atmospheric $H$ maps are $<1$, supporting the view that winter SSTA behaves as an integrated response to atmospheric fluctuations; in other words, the OU-like model is an appropriate abstraction for the winter EJS. Besides the basin-scale coherence of SLP and 10m winds, the CurlTau map shows frontal-parallel patchiness, foreshadowing the along-front belts recovered by the MCA. These patterns motivate a coupled, multivariate analysis.

FIGURE 2 TO BE INSERTED

### 3.2 SSTA co-variability and its atmospheric counterparts in AO-conditioned winters

Because winter forcing exhibits strong synoptic scales, we use a rank-reduced, multivariate MCA to extract coupled modes between SSTA and the concatenated atmospheric fields. Unless otherwise noted, all 1$^{st}$-mode MCA-EOF loadings are unitless spatial weights computed separately for +AO and −AO; Figures 3–4 show masked versions that retain salient grid cells (Eq. 16), and Figure S2 provides unmasked fields for reference. This design allows us to isolate AO-phase-dependent hot spots (EKB, SPF) and to relate SSTA to co-varying SLP/wind/curl structures that will be interpreted with the OU-based time-scale diagnostics below.

#### 3.2.1 Wintertime SSTA 1$^{st}$-mode loading

The first MCA-EOF mode of winter SSTA exhibits positive loadings concentrated over the East Korean Bay and along a NW–SE belt near 40–41°N in both AO phases, with localized maxima around 38–40°N, 129–132°E and 40–41°N, 133–137°E (Figure 3). In −AO winters the near-coastal maximum intensifies and extends westward, while in +AO winters the offshore lobe is relatively stronger. The band-wise geometry broadly follows the climatological subpolar front near ~40°N and the region of largest winter SST variance in the NW EJS, consistent with prior basin structures (Song et al. 2023). Also, the unmasked SSTA loadings (Figure S1) show that this front-parallel structure is spatially continuous and not an artifact of masking. The pattern, by construction, reflects covariance with the concurrent atmospheric fields, motivating the joint interpretation below.

FIGURE 3 TO BE INSERTED

#### 3.2.2 2m air temperature anomaly (ATMPA) 1$^{st}$-mode loading

ATMPA loadings are basin-wide positive in both AO phases, with broader south–westward coverage in −AO winters (Figure 4A, B). Unmasked panels emphasize a smooth meridional gradient across the basin (Figure S2A, B). Coherent ATMPA-SSTA loadings indicate co-occurrence of warm near-surface air and warm SST over the NW EJS, yet the air-sea heat flux anomalies alone do not reproduce the NW-corner SST pattern, implying a significant role for ocean dynamic adjustments (Song et al. 2023).

FIGURE 4A, B TO BE INSERTED

### 3.2.3 Wind-stress curl anomaly (CurlTauA) 1$^{st}$-mode loading

In +AO winters, CurlTauA loadings form alternating lobes: like-signed cells over the East Korean Bay (≈36–39°N, 129–133°E), an oppositely signed offshore belt oriented SW–NE (≈39–41°N, 135–140°E), and additional features north of 42°N (Figure 4C). In −AO winters, a quasi-continuous belt of like-signed loadings aligns along 40–42°N, 130–134°E from the Vladivostok coast toward the North Korean margin (Figure 4D). Unmasked maps highlight the continuity of this along-front belt and its SE counter-lobe near 136–140°E (Figure S2D)

Mechanically, these co-varying curl patterns are precisely where Song et al. (2023) documented AO-linked Ekman pumping and eddy/SSH responses: during +AO, anomalous anticyclonic curl and Ekman downwelling over the NW EJS co-occur with positive SSH and anticyclonic eddy-like circulation, favoring local SST warming; the opposite holds in −AO. The front-parallel belt in our −AO loading (40–42°N) lies on the climatological subpolar-front latitude and where winter variance is largest, consistent with front-following wind-curl anomalies that can modulate vertical motion and eddy activity along the front.

FIGURE 4C, D TO BE INSERTED

### 3.2.4 Sea-level pressure anomaly (SLPA) 1$^{st}$-mode loading

SLPA loadings show a north-south gradient: positive over the far-northern sector in +AO and negative basin-wide south of ~41–42°N in −AO (Figure 4E, F; Figure S2E, F). This configuration is consistent with the AO-conditioned lower-tropospheric circulation over the EJS reported by Song et al. (2023; their Figure 5) – namely, weaker north-westerlies in +AO and stronger north-westerlies in −AO, which set the sign and structure of the surface wind and curl anomalies above.

FIGURE 4E, F TO BE INSERTED

### 3.2.5 10m zonal wind anomaly (UA10) and meridional wind anomaly (VA10) 1$^{st}$-mode loading

UA10 loadings are predominantly of one sign across the basin in both phases (more spatially coherent in +AO; Figure 4G, H), while VA10 loadings are positive over the central-northern basin with a SW–NE tilt (Figure 4I, J). Unmasked panels clarify the basin-scale coherence and reveal secondary tongues near the NW shelf that are partly truncated by the mask (Figure S2G, H). Taken together with SLPA, these wind patterns provide the near-surface forcing that co-varies with CurlTauA, aligning with the "AO-modulated wind-stress-curl → Ekman pumping pathway" documented by Song et al. (2023; their Figures 5–8).

FIGURE 4G, H TO BE INSERTED

FIGURE 4I, J TO BE INSERTED

### 3.2.6 Cross-field consistency with ocean dynamic adjustment (link to SSTA mode)

The spatial co-location of (i) positive SSTA loadings over the East Korean Bay and along ~40°N, (ii) front-parallel CurlTauA belts, and (iii) AO-consistent SLP/wind anomaly loadings indicates that the leading coupled mode over the NW EJS is dynamically tied to the AO-forced surface wind field. This is consistent with Figures 6–8 of Song et al. (2023), who showed that net heat flux (NHF) anomalies do not reproduce the NW-corner SST pattern, whereas curl-driven Ekman downwelling/upwelling and eddy-like SSH responses do. Moreover, Figures 10–11 of Song et al. (2023) demonstrated that the resulting warm/cold anomalies extend through the upper ~100–200m and propagate seasonally into the subsurface, a vertical signature anticipated for the front-locked structures we recover here.

Our AO-conditioned loadings mirror the observed phase dependence of winter SST/MHW over the NW EJS: +AO winters favor anticyclonic eddy-like circulation, downwelling, increased SSH, and abnormally warm SSTs, conducive to more MHW days; −AO winters favor the converse (See Song et al., 2023, Figures 8,12,13).

## 3.3 Response timescales and persistence of the coupled 1st-mode

### 3.3.1 Zero-lag cross-correlations: SSTA versus OU-integrated atmospheric field

For each atmospheric field's 1st-mode PC, we constructed an OU-like integrated response series with integration time step $\tau$ (days). We then computed zero-lag correlations with the 1st-mode SSTA PC for +AO and −AO winters separately. The optimal integration time $\tau^*$ is defined as the $\tau$ at which correlation either peaks or begins a clear plateau.

FIGURE 5 TO BE INSERTED

- **ATMPA (2m air temperature)**: Correlations rise rapidly and plateau at $\tau^* \approx$ 20–30 days, reaching $r \sim$ 0.65–0.70 in both AO phases; the plateau is slightly broader under −AO. This indicates that mixed-layer thermal memory of ~3–4 weeks efficiently transmits near-surface air temperature anomalies into SSTA. Consistent with Song et al. (2023), however, net surface heat flux patterns alone do not reproduce the NW-corner SST anomaly, implying that ATMPA contributes in tandem with ocean dynamics.
- **CurlTauA (wind-stress curl)**: Correlations climb to $r \sim$ 0.58–0.62 with a broad plateau at $\tau^* \approx$ 18–25 days in both AO phases. The narrow $\tau^*$ range and high $r$ are consistent with a fast dynamical pathway; curl anomalies force Ekman pumping, which in turn organizes eddy/SSH anomalies that warm/cool the NW EJS mixed layer (anticyclonic/downwelling in +AO, cyclonic/upwelling in −AO).
- **SLPA (sea-level pressure)**: Correlations increase more gradually, peaking near the end of the tested window; $\tau^* \approx$ 40–50 days, with stronger values in −AO ($r \sim$ 0.40–0.45) than in +AO ($r \sim$ 0.30–0.35). This suggests that large-scale pressure patterns (AO-like) project onto the EJS surface winds persistently on 5–7-week timescales in −AO winters, coherently preconditioning the curl field over the basin, consistent with the AO-conditioned 850-hPa composites in Song et al. (2023)

- **UA10 (10m zonal wind)**: Correlations peak at $\tau^* \approx$ 15–20 days with $r \sim$ 0.35–0.40, tending to be higher and more sustained in +AO. This supports the view that +AO winters feature zonal-wind anomalies more directly with the NW-corner anticyclonic/SSH response, in line with the +AO composites of curl and SSH.
- **VA10 (10m meridional wind):** Correlations grow longest for meridional wind in −AO, reaching $r \sim$ 0.50–0.55 with $\tau^* \approx$ 30–40 days; in +AO, values are modest ($r \sim$ 0.20–0.30). This AO-phase contrast is consistent with stronger north-westerlies in −AO (Song et al. 2023; 850-hPa composites), which reinforce the front-parallel curl belts and the associated Ekman pumping near ~40–42°N.

Across variables, $\tau^*$ clusters into ~3 weeks for curl/ATMPA/UA10 and ~4–7 weeks for SLPA/VA10 (−AO). This hierarchy maps cleanly onto the AO-forced mechanism over the NW EJS: pressure → wind (zonal/meridional) → curl → Ekman pumping → eddy/SSH → SST, with the curl/Ekman/eddy segment responding on ~2–4 weeks and the large-scale pressure/meridional-wind conditioning on ~1–2 months. The amplitude and phasing difference between +AO and −AO mirror Song et al.'s finding that +AO favors anticyclonic/downwelling/SSH↑/warm SST and more MHW days, whereas −AO favors cyclonic/upwelling/SSH↓/cool SST.

### 3.3.2 Persistence diagnostics from DFA

To assess whether our OU-integrated atmospheric PCs attain persistence properties comparable to SSTA (as expected from Hasselmann's stochastic climate framework), we applied DFA to the SSTA PC and to each integrated atmospheric PC (here with $\Delta\tau = 25$ days).

- The SSTA Hurst exponents are $\approx$ 1.40 in +AO and $\approx$ 1.31 in −AO, indicating a reddened, highly persistent process.
- The integrated fields exhibit comparable exponents: ATMPA$_{int}$ ~ 1.39(+AO) / 1.38(−AO); CurlTauA$_{int}$ ~ 1.31(+AO) / 1.27(−AO); SLPA$_{int}$ ~ 1.31 / 1.22; UA10$_{int}$ ~ 1.29 / 1.35; VA10$_{int}$ ~ 1.25 / 1.05 (values from panel insets of Figure 6).

FIGURE 6 TO BE INSERTED

These DFA results confirm that our OU-like transform yields forcing series with SSTA-like memory, lending methodological validity to using $\tau^*$ (estimated from Figure 5) as process-based predictors in subsequent statistical/dynamical prediction experiments.

### 3.3.3 Why integration matters: comparison with raw lead-lag correlations

As shown in Figure S3, lead-lag cross-correlations between raw atmospheric PCs (forcing leading) and the SSTA PC are weak ($|r| \lesssim$ 0.1–0.2 near short lags) and often change sign at multi-week lags across variables and AO phases. In contrast, OU-integrated series (Figure 5) yield substantially larger zero-lag correlations (typically $r \sim$ 0.4–0.7) with clear $\tau^*$ plateaus. This contrast demonstrates that mixed-layer integration is essential to capture how atmospheric variability imprints on winter SSTA in the NW EJS—precisely the mechanism advocated by Song et al. (2023), where AO-related wind forcing modulates SST via Ekman pumping and eddy/SSH adjustment rather than by local heat flux alone.

### 3.3.4 AO-phase dependence and forecasting value

Together, Figures 5–6 indicate phase-specific sensitivities:

- In +AO, UA10 and CurlTauA with $\tau^* \sim$ 2–3 weeks are the most efficient SSTA covariates, consistent with anticyclonic/downwelling/SSH ↑ responses near the East Korean Bay and along the western Japan margin.
- In −AO, VA10 and SLPA require longer integration ($\tau^* \sim$ 4–7 weeks), consistent with more persistent NW-monsoonal flow and basin-scale pressure control that organize front-parallel curl belts near 40–42°N, while CurlTauA remains a robust 2–3-week predictor of the direct dynamical response.

These results provide process-informed predictor windows for winter extremes (e.g., MHWs or cold surges impacting SST); (1) use 3-week OU-integrated CurlTauA/ATMPA/UA10, (2) augment with 4–7-week OU-integrated VA10/SLPA under −AO, and (3) condition the regression on AO phase. This aligns with the observation that +AO winters show more MHW days (≈12.5 vs 0.5 in −AO) in the NW EJS through the curl-driven dynamical pathway (Song et al. 2023).

## 4 Discussion

### 4.1 AO-phase control of coupling geometry and memory

Our diagnostics resolve how AO polarity structures the full winter pathway from pressure → near-surface wind → wind-stress curl → Ekman pumping → eddy/SSH → SSTA in the NW EJS. The persistence maps establish that SSTA is reddened and highly persistent ($H \approx$ 1.3–1.5), while atmospheric drivers are less persistent ($H < 1$), consistent with a Hasselmann-type integrator. The MCA then shows that the leading coupled mode (SCF $\approx$ 0.87 in +AO; $\approx$ 0.75 in −AO) locks onto hot spots over EKB and along the SPF, where CurlTau organizes front-parallel belts whose sign and placement differ by AO phase. Finally, OU-integrated correlations quantify effective memory times: ~ 18–25 days for CurlTau (fast dynamical pathway), ~ 15–30 days for ATMP/U10, and ~ 30–50 days for SLP/V10 (slow, basin-scale conditioning), with longer $\tau^*$ for SLP and V-wind in −AO. Together, these lines of evidence indicate that +AO winters emphasize a short-memory, curl-driven warm response in the NW corner, whereas −AO winters emphasize longer-memory pressure/meridional-wind control that organizes along-front curl belts near 40–42°N.

### 4.2 Why integration is essential for subseasonal predictability

Lead-lag correlations between raw atmospheric PCs and the SSTA PC are weak and sign-changing at multi-week lags, whereas OU-integrated forcings yield strong zero-lag correlations with clear plateaus. Physically, this reflects the mixed-layer integration of high-frequency weather into ocean memory, which reddens the SST spectrum and elevates persistence. The plateau behavior in $\tau$ implies diminishing returns beyond ~ 3–7 weeks, and the AO-phase contrast indicates that the memory kernel itself is state-dependent. From a prediction standpoint, the OU transform supplies process-aware predictors with tunable windows matched to the integration physics rather than to arbitrary lags.

### 4.3 Implications for extremes and phase-aware windows

Song-type composites relate +AO to anticyclonic curl, downwelling, SSH ↑, and warm SSTA/MHWs, with the converse in −AO (Song et al., 2023). Our time-scale results explain why these composites materialize: in +AO, short-memory drivers (CurlTau/ATMP/U10, $\tau^*$ ~ 2–3 weeks) project efficiently onto the EKB-SPF corridor, sustaining warm anomalies and enhancing MHW propensity; in −AO, longer-memory SLP/V10 ($\tau^*$ ~ 4–7 weeks) precondition the basin and organize curl belts that favor cool anomalies and cold-surge impacts on SST. These phase-specific predictor windows–(1) 3-week OU-integrated CurlTau/ATMP/U10 for all winters, (2) add 4–7-week OU-integrated SLP/V10 in −AO– offer concrete lead times for probabilistic extremes prediction using OU-driven ensembles of the SSTA PC (and mapped back to space via MCA loadings).

### 4.4 Limitation and outlook

Our analysis adopts a linear, single-time-scale OU kernel and isolates one leading coupled mode. While this suffices to expose clear AO-dependent memory windows and validates the Hasselmann integrator view, it does not resolve potential multi-time-scale kernels, nonlinearity with background state, or mesoscale-synoptic interactions. These aspects will be explored in follow-ups by (i) allowing state-dependent $\tau$ (and multiplicative noise) within the OU framework, (ii) testing multi-mode MCA reconstructions, and (iii) embedding the identified $\tau^*$ predictors into cross-validated hindcast experiments for winter extremes (MHWs and cold surges) in the EJS.

### 4.5 Summary

Winter SSTA persistence in the EJS is consistent with a Hasselmann integrator, and AO polarity reorganizes both the geometry (EKB/SPF hot spots; curl belts) and the memory (2–4 weeks for curl/ATMP/U10; 4–7 weeks for SLP/V10) of air-sea coupling. These findings provide phase-aware, process-based predictor windows that can initialize OU-based ensembles for subseasonal extremes– marine heatwaves and cold-surge-impacted SST– over the NW EJS.

## 5 Conflict of Interest Statement

The authors declare that the research was conducted in the absence of any commercial or financial relationships that could be considered as potential conflicts of interest.

## 6 Data Availability Statement

The monthly AO index was obtained from the NOAA Climate Prediction Center (CPC) website at https://www.cpc.ncep.noaa.gov/ (accessed on 10 April 2024).

The daily NOAA OI SST V2 High Resolution Dataset data used in this study is publicly available from the NOAA Physical Sciences Laboratory (PSL) at https://psl.noaa.gov/data/gridded/data.noaa.oisst.v2.highres.html (accessed on 4 April 2023).

The ERA5 reanalysis data, including 10m U and V wind components, sea level pressure, and 2m atmospheric temperature are publicly available from the Copernicus Climate Data Store

(CDS) at https://cds.climate.copernicus.eu/cdsapp#!/dataset/reanalysis-era5-single-levels (accessed on 10 April 2025).

# 7 Author Contributions

Conceptualization, G.L. and J.-J.P.; methodology, G.L; formal analysis, G.L.; investigation, G.L.; data curation, G.L.; writing—original draft preparation, G.L.; writing—review and editing, G.L. and J.-J.P.; visualization, G.L.; supervision, J.-J.P.; funding acquisition, G.L. and J.-J.P. All authors have read and agreed to the published version of the manuscript.

# 8 Acknowledgements

# 9 Funding

This research was supported by the Korea Institute of Marine Science & Technology Promotion (KIMST) funded by the Ministry of Oceans and Fisheries (RS-2023-00256005). This work was also supported by the National Research Foundation of Korea (NRF) grant funded by the Korean government (MSIT) (RS-2022-NR069134). G.L. was supported by the National Research Foundation of Korea (NRF) grant funded by the Korean government (MSIT) (RS-2024-00507484).

## 11    Figure Legends

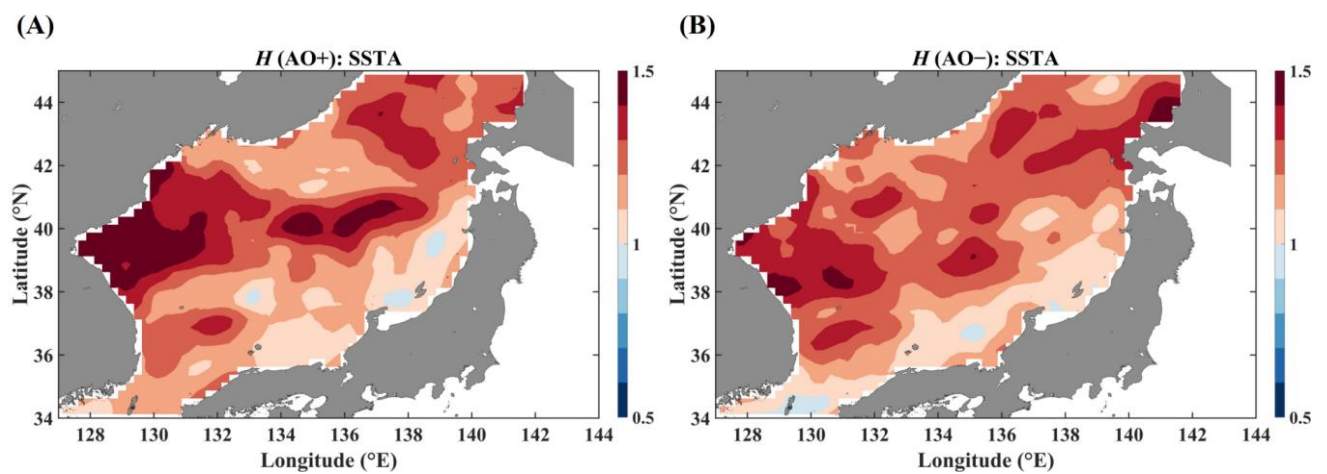

**FIGURE 1**. Spatial distribution of the Hurst exponent *H* for wintertime (JFM) SST anomalies (SSTA) over the East/Japan Sea. Panels show (A) positive-AO winters and (B) negative-AO winters. *H* was estimated with DFA over 10–90-day scales (Section 2.2.1). Values *H*>1 indicate random-walk-like persistence consistent with mixed-layer integration of weather noise. Notable hot spots occur in East Korea Bay and along the Sub-Polar Front (~40–41°N), with higher persistence under +AO.

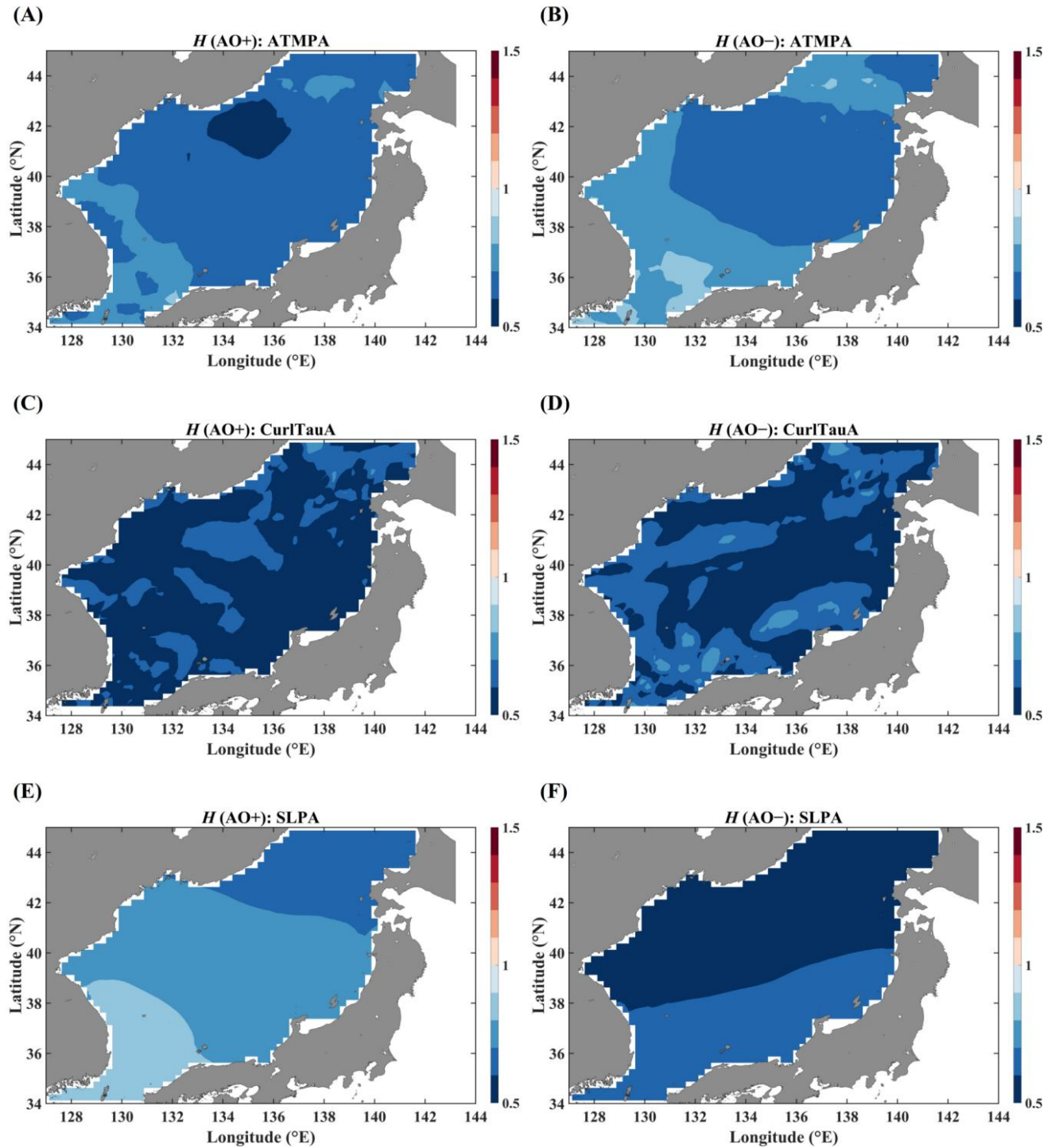

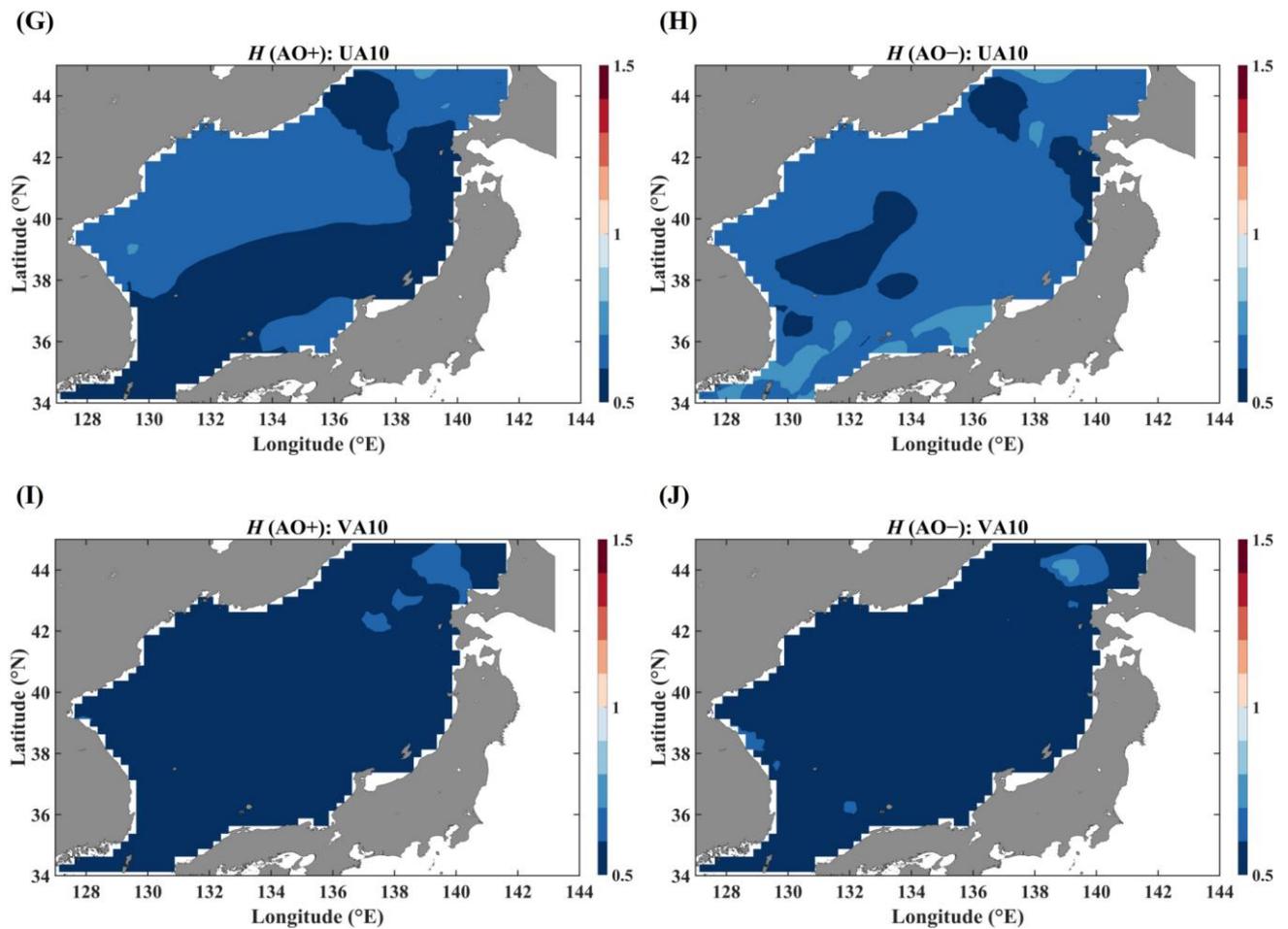

**FIGURE 2**. Hurst exponent $H$ maps (10–90-day scales) for (A,B) 2m air temperature anomaly (ATMPA), (C,D) wind-stress-curl anomaly (CurlTauA), (E,F) sea-level-pressure anomaly (SLPA), (G,H) 10m zonal wind anomaly (UA10), and (I,J) 10m meridional wind anomaly (VA10), for +AO and −AO winters respectively. Atmospheric fields generally exhibit $H<1$, consistent with their role as stochastic forcings; CurlTauA shows front-parallel patchiness near 40–42°N (−AO), indicative of localized Ekman-pumping control.

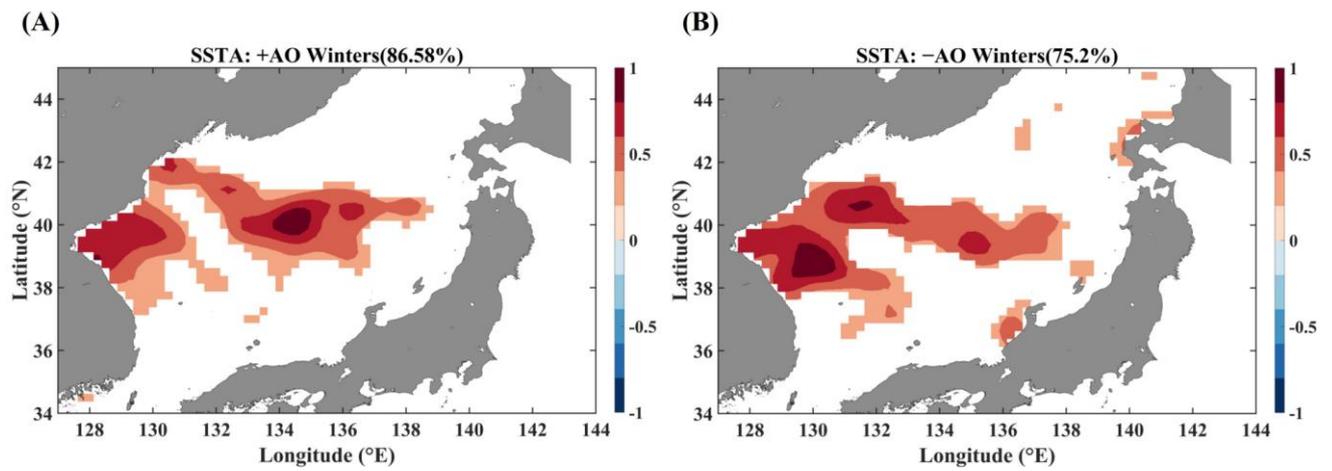

**FIGURE 3**. Unit-variance, max-absolute-normalized loadings of the 1st MCA mode for SSTA in (A) +AO and (B) −AO winters (masked by the area-weighted saliency criterion; Section 2.2.3). Positive loadings align with East Korea Bay and a NW-SE belt near the Sub-Polar Front. Unmasked fields are provided in Supplementary Figure S1.

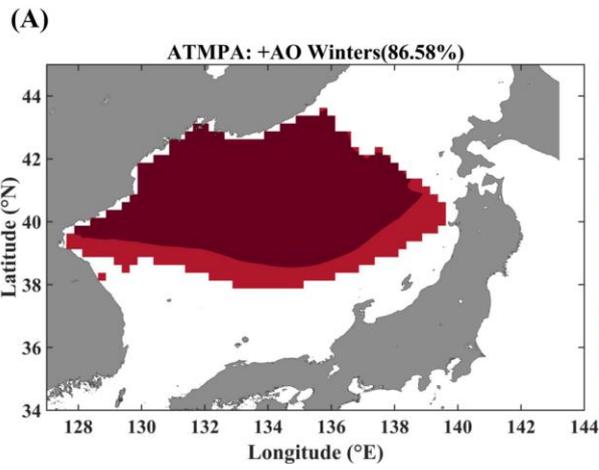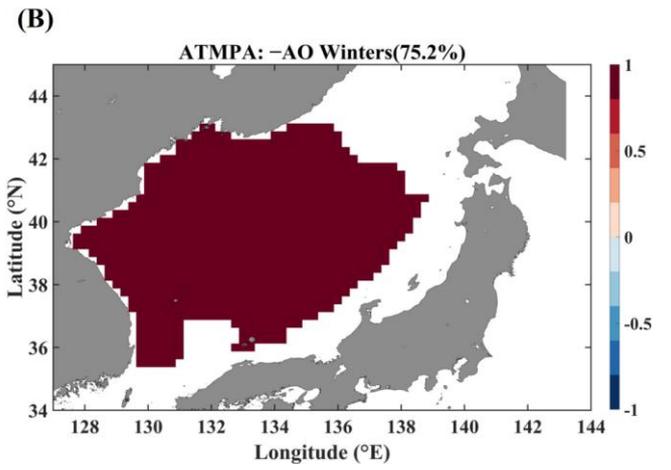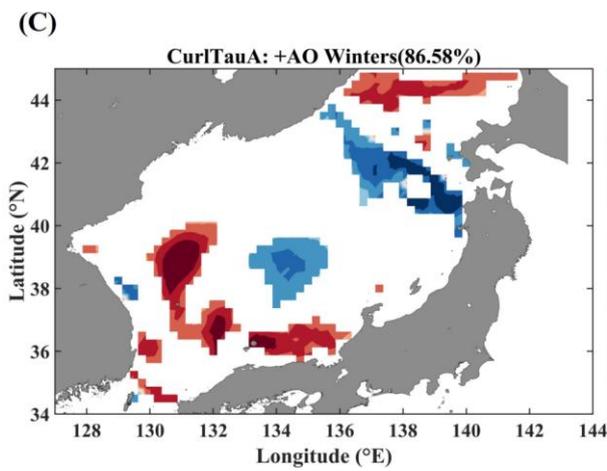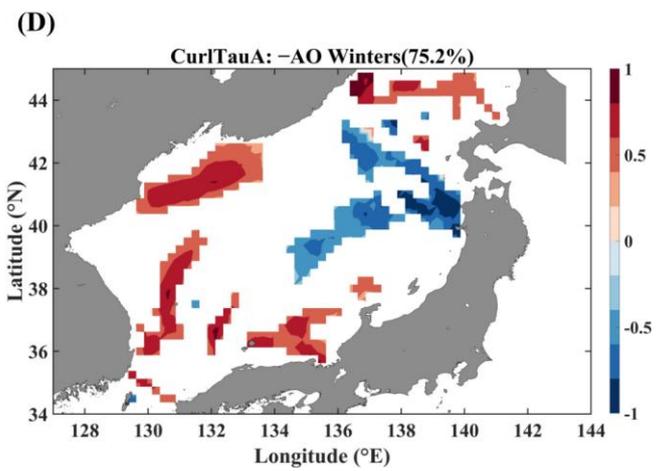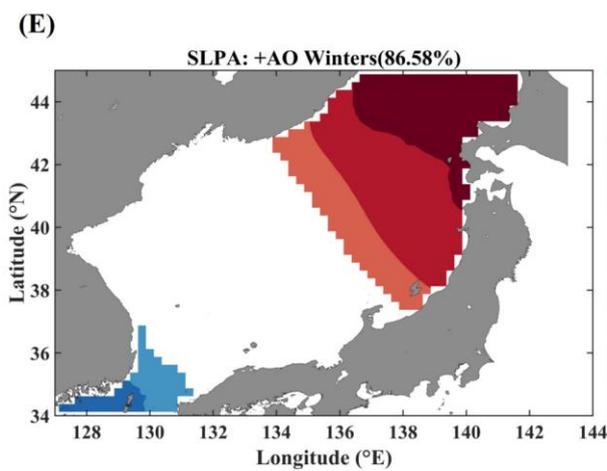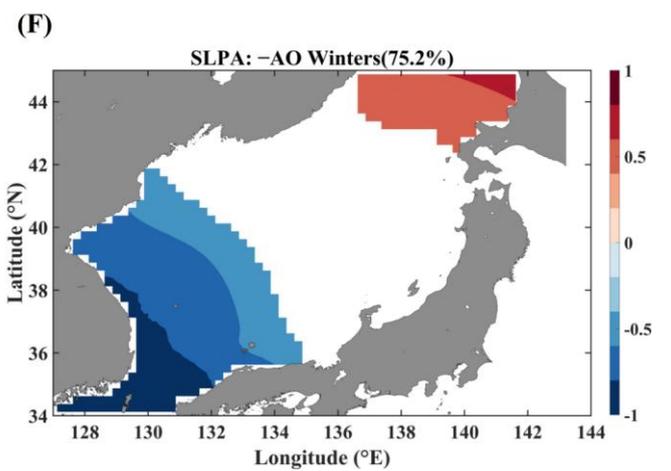

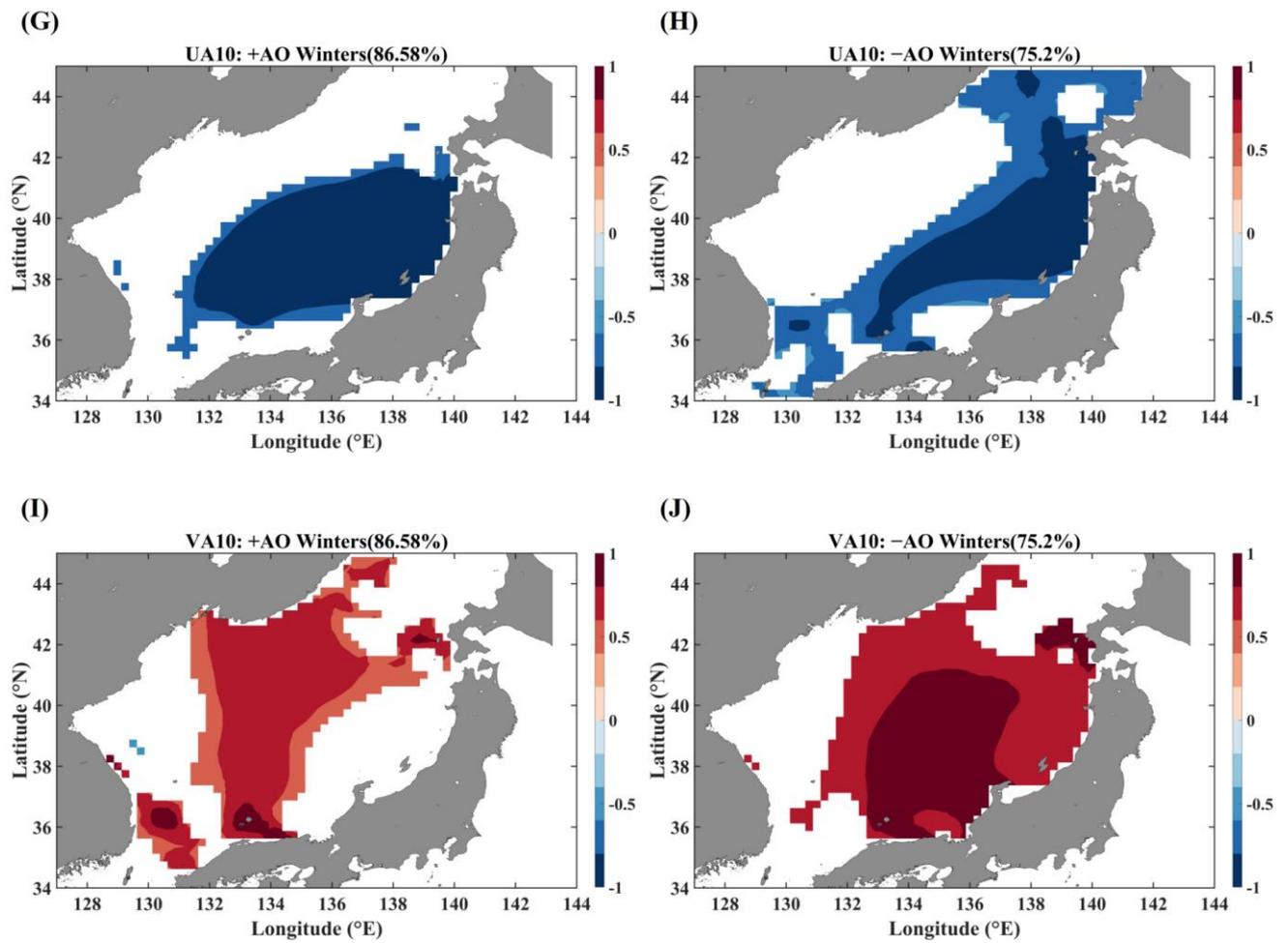

**FIGURE 4**. Same format as Figure 3, showing paired atmospheric loadings for (A,B) ATMPA, (C,D) CurlTauA, (E,F) SLPA, (G,H) UA10, and (I,J) VA10 in +AO and −AO winters. Saliency-masked maps emphasize robust structures; basin-scale ATMPA/SLPA/UA10/VA10 patterns and a front-parallel CurlTauA belt (~40–42°N) that is prominent in −AO winters. Full, unmasked fields appear in Supplementary Figure S2.

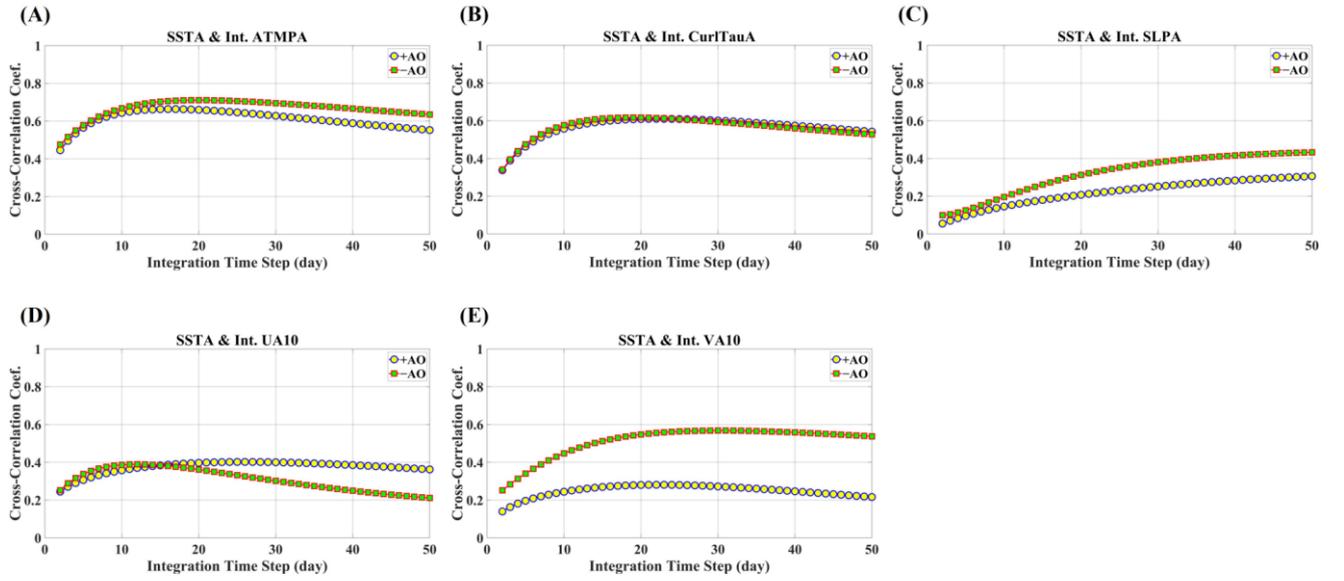

**FIGURE 5**. For each atmospheric variable, the 1st-mode atmospheric PC was convolved with an exponential kernel of e-folding $\tau$ (2–50 days) to form an integrated response series (Section 2.2.4). Curves show the zero-lag Pearson correlation with the SSTA 1st PC for +AO (blue) and −AO (red). Optimal integration time $\tau^*$ is defined by the peak or onset of a plateau. Across variables, $\tau^*$ clusters near 2–3 weeks for ATMPA, CurlTauA, and UA10, and 4–7 weeks for SLPA and VA10 (especially in −AO).

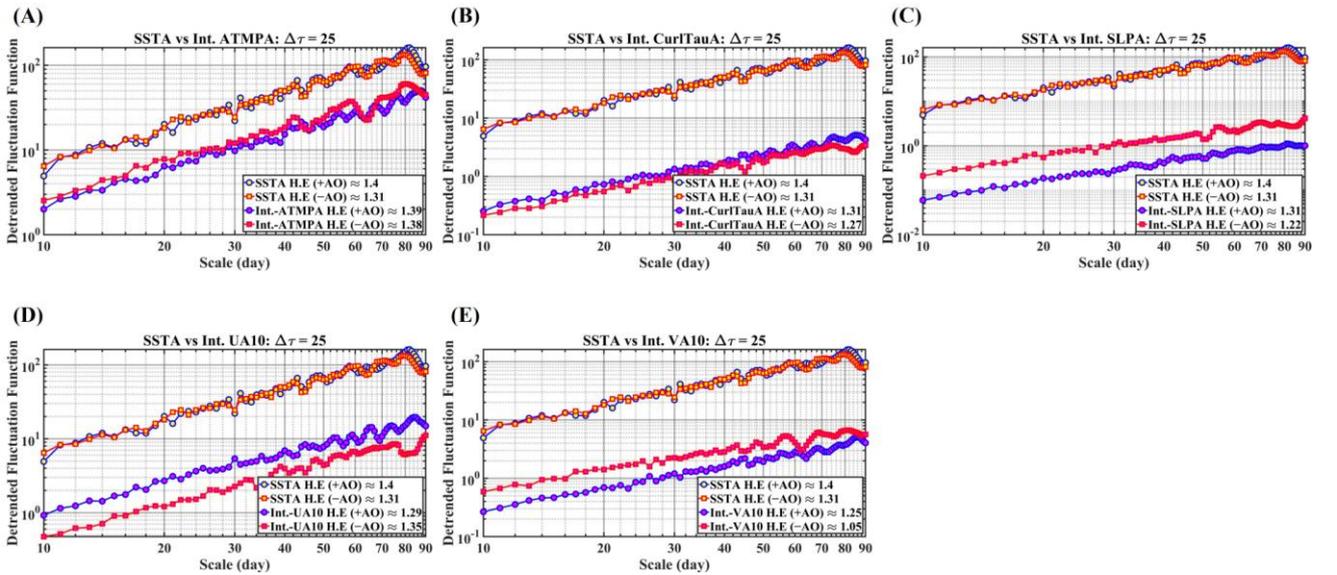

**FIGURE 6**. Detrended Fluctuation Analysis of the SSTA first PC and the OU-integrated atmospheric PCs for each field. Slopes of the log-log fluctuation functions provide **H**; panel insets list **H** for +AO and −AO. Integrated responses attain ocean-like persistence (**H**≈1.0–1.5), validating the Hasselmann/OU interpretation adopted here.

## *Supplementary Material*

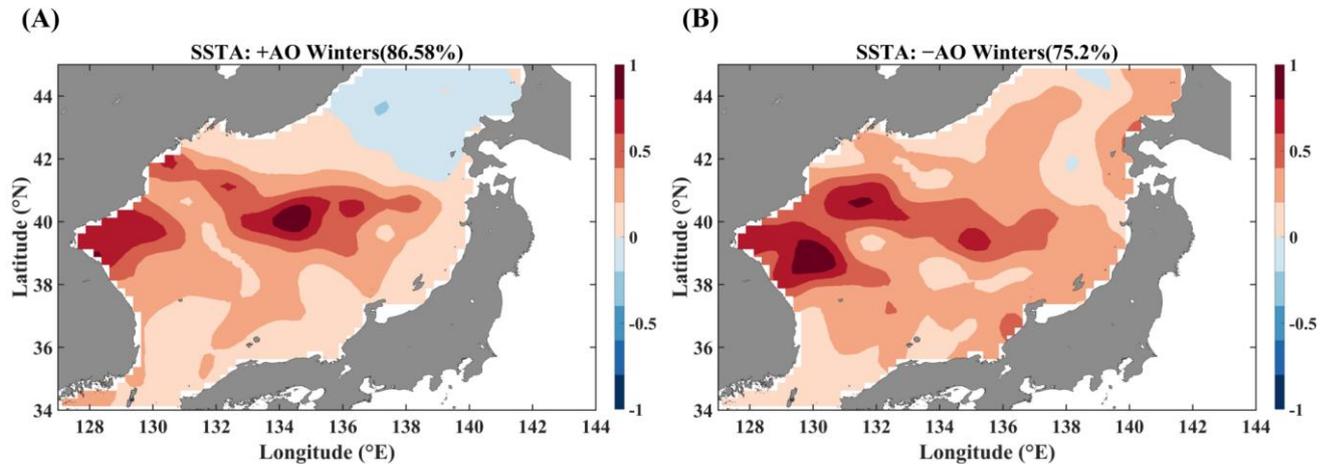

**Supplementary Figure S1.** As Figure 3, but without the saliency mask; fields are spatially continuous along the Sub-Polar Front and in East Korea Bay.

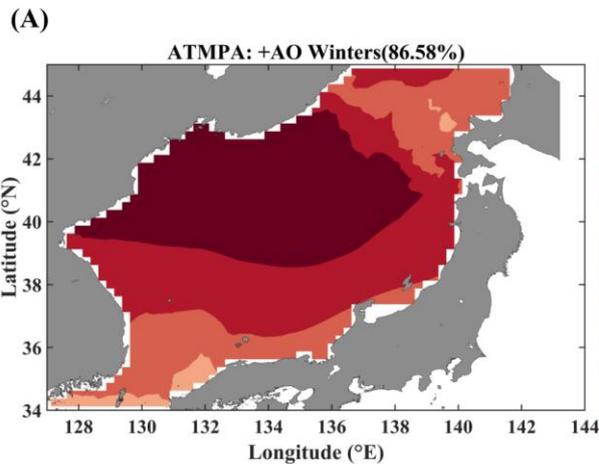
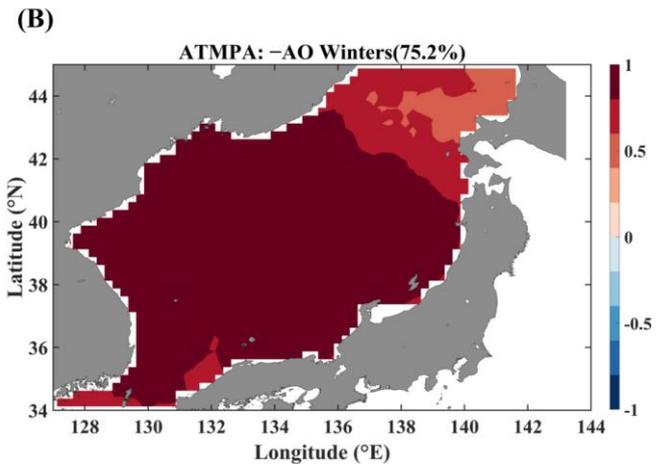
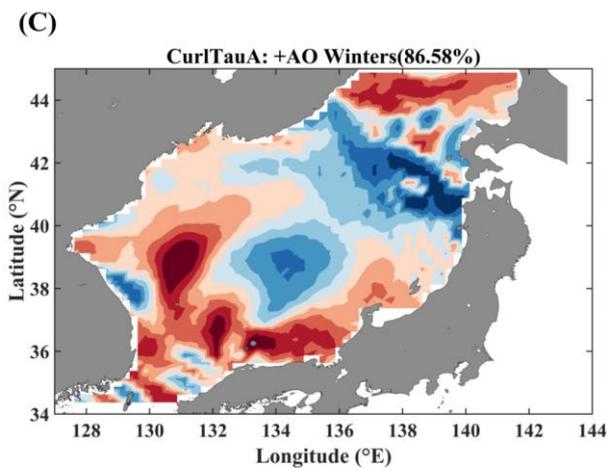
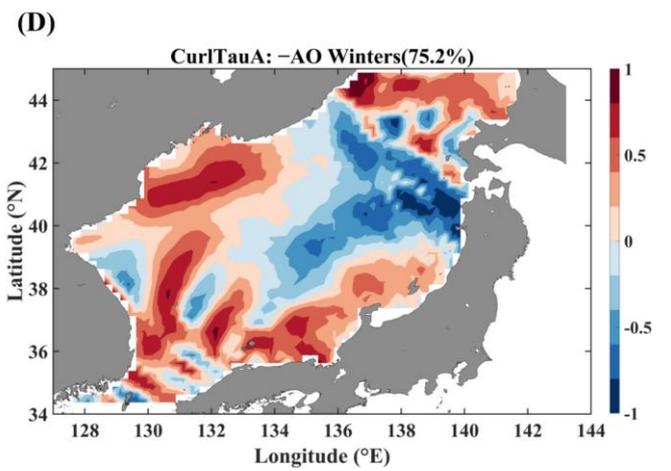
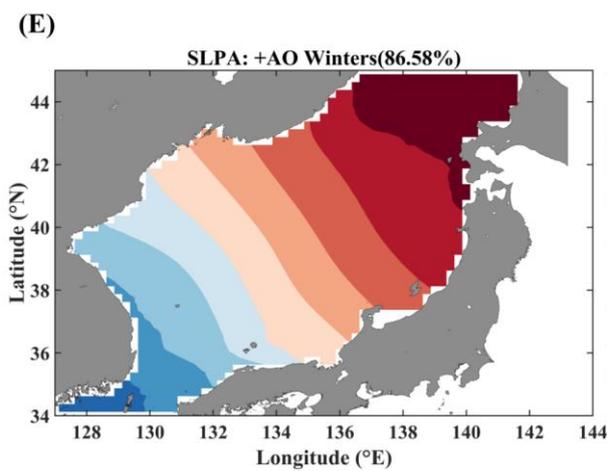
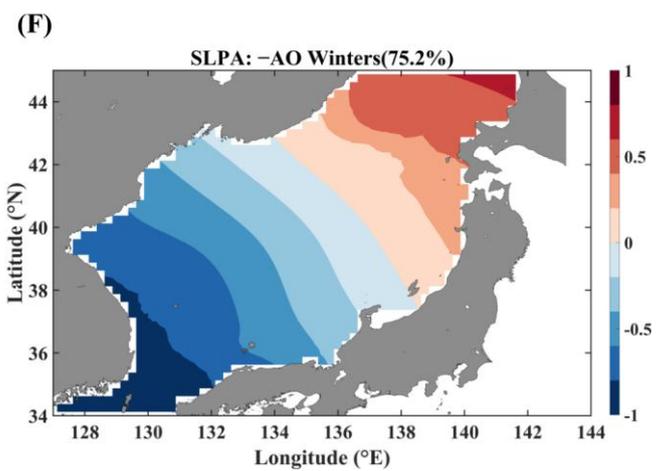

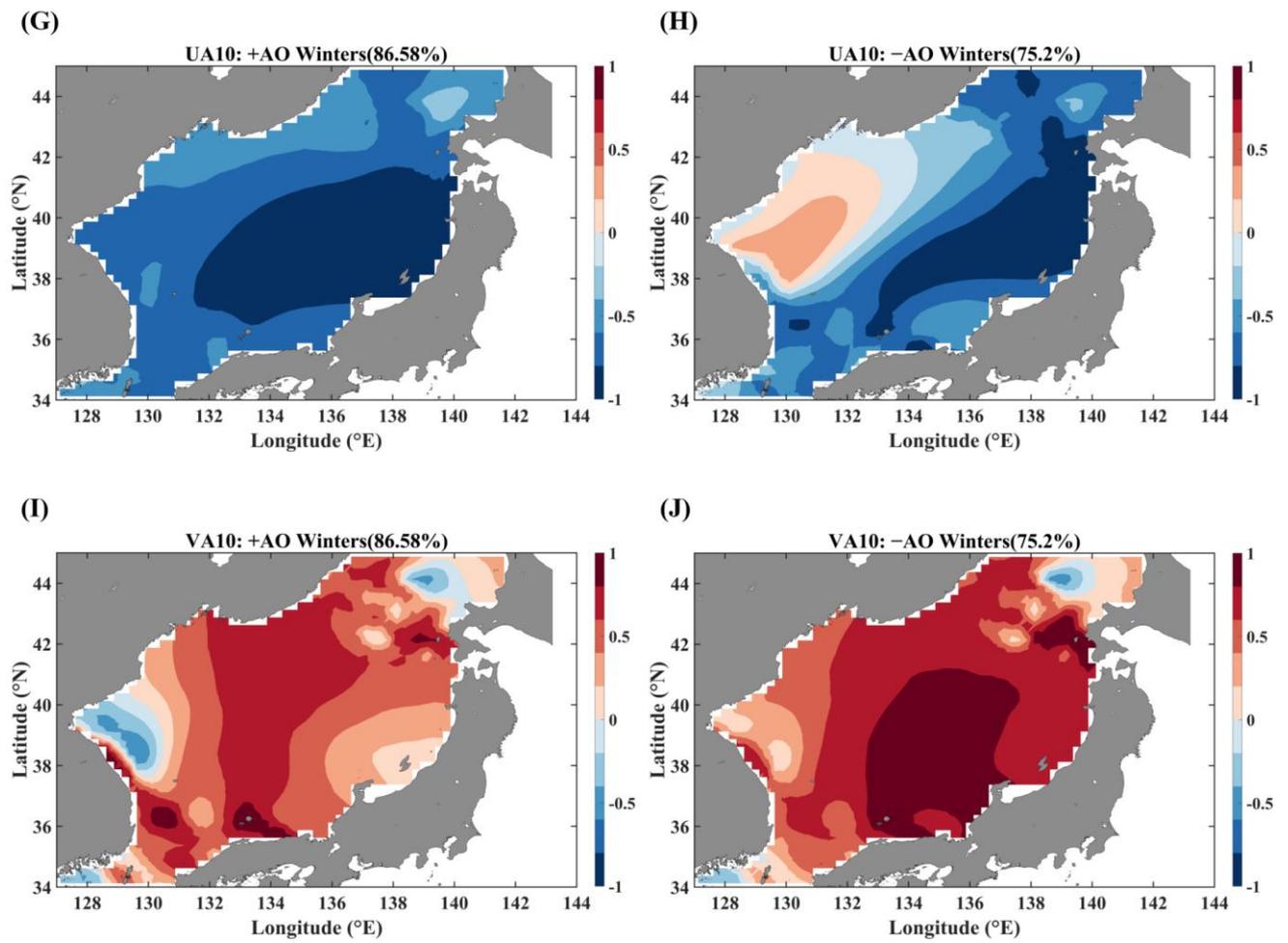

**Supplementary Figure S2**. As Figure 4, but without masking. The −AO CurlTauA field exhibits a quasi-continuous belt along 40–42°N from the Vladivostok coast toward the North Korean margin.

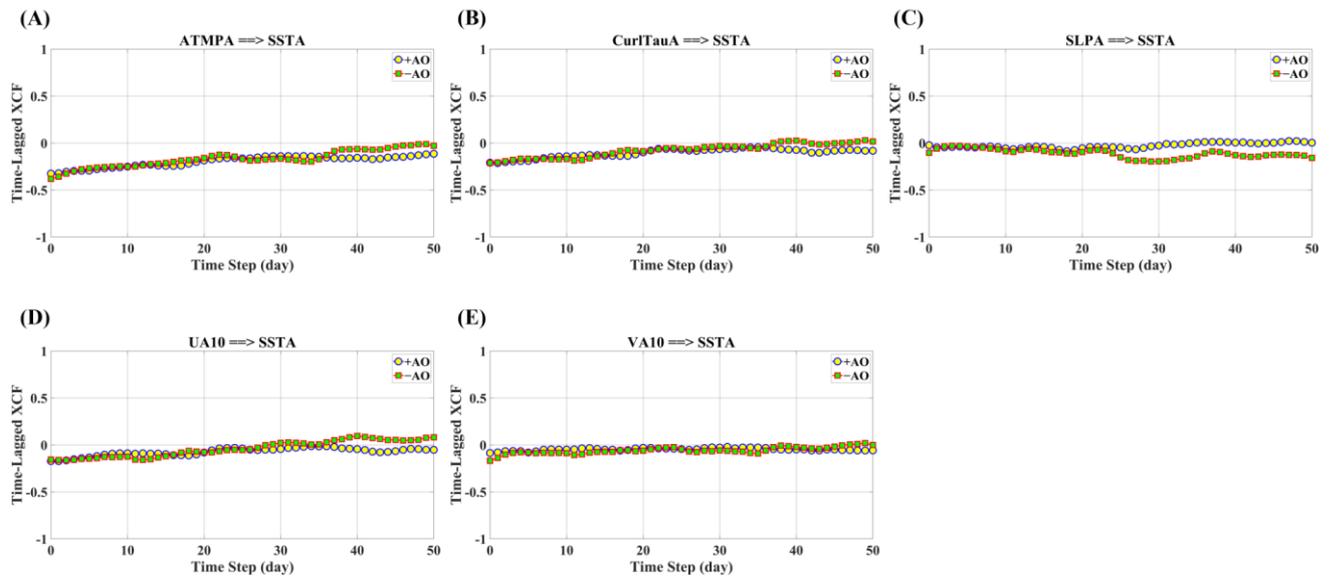

**Supplementary Figure S3**. Time-lagged cross-correlations between raw atmospheric first PCs and the SSTA first PC for +AO (blue) and −AO (red). Correlations are weak and sign-variable compared with Figure 5, emphasizing the need for mixed-layer integration (OU-type) when diagnosing air-sea coupling.